\documentclass[aps,pre,twocolumn,showpacs,floatfix,superscriptaddress]{revtex4}
\usepackage{graphicx}
\usepackage{dcolumn}
\usepackage{bm}
\usepackage{epsf}
\usepackage{subfigure}
\usepackage{epstopdf}
\usepackage{amsmath}
\usepackage{amssymb}
\usepackage[cspex,bbgreekl]{mathbbol}
\usepackage{color}

\newcommand{\E}{{\mathcal E}}
\newcommand{\A}{{\mathcal A}}

\usepackage{braket}

\begin{document}

\title{Nonequilibrium potential and fluctuation theorems for quantum maps}

\author{Gonzalo Manzano}\affiliation{Departamento de F\'isica At\'omica, Molecular y Nuclear and  GISC, Universidad Complutense Madrid, 28040 Madrid, Spain}
\affiliation{Instituto de F\'isica Interdisciplinar y Sistemas Complejos IFISC (CSIC-UIB),
Campus Universitat Illes Balears, E-07122 Palma de Mallorca, Spain}
\author{Jordan M. Horowitz}
\affiliation{Department of Physics, University of Massachusetts at Boston, Boston, MA 02125, USA}
\author{Juan MR Parrondo}
\affiliation{Departamento de F\'isica At\'omica, Molecular y Nuclear and  GISC, Universidad Complutense Madrid, 28040 Madrid, Spain}

\date{\today}

\begin{abstract}
We derive a general fluctuation theorem for quantum maps. The theorem applies to a broad class of quantum dynamics, such as unitary evolution, decoherence, 
thermalization, and other types of evolution for quantum open systems. The theorem reproduces well-known fluctuation theorems in a single and simplified 
framework and extends the Hatano-Sasa theorem to quantum nonequilibrium processes. Moreover, it helps to elucidate the physical nature of the environment inducing a given dynamics in an open quantum system.
\end{abstract}

\pacs{
05.70.Ln,  
05.40.-a   
05.70.-a   
}
\maketitle

\section{Introduction}

Completely-positive, trace-preserving (CPTP) quantum maps capture a vast diversity of quantum dynamical evolutions, including arbitrary open-system dynamics such as decoherence, measurement, and thermal relaxation \cite{kraus,breuer}. Consequently, the thermodynamic analysis of processes described by CPTP maps is a major issue in the development of quantum thermodynamics~\cite{Yukawa:2001tf,Sagawa:2012vd,Horowitz2014b,Goold:2014vx,Binder2015}. One of the main tools of such thermodynamic analysis are fluctuation theorems, since they reveal the statistical properties of thermodynamic quantities such as work or entropy production along arbitrary nonequilibrium processes \cite{Campisi:2011ka,Esposito2009,Deffner2011}. Therefore, fluctuation theorems for arbitrary CPTP maps will be relevant to understand the role of quantum effects in thermodynamics. 

In recent years, there have been several derivations of fluctuations theorems for specific classes of CPTP maps falling into two broad categories: detailed fluctuation theorems for quantum trajectories and fluctuation theorems for thermodynamic variables, such as work and entropy.
Campisi~\emph{et al.}~obtained a detailed fluctuation theorem for a unitary, driven evolution punctuated by unital maps -- maps for which the identity matrix is invariant, such as projective measurements --~\cite{Campisi:2010kx,Campisi:2011gp}.  This work was followed up and extended by Watanabe~\emph{et al.}~\cite{Watanabe:2014fh}.
General quantum Markov semigroups were explored by Crooks using time-reversed or dual maps~\cite{Crooks:2008ji}, which was then applied by Horowitz \emph{et al.}\ to nonequilibrium quantum jump trajectories
~\cite{Horowitz:2013jd,Horowitz2014b}.
An alternative, operator formulation for driven Lindbald master equations was independently developed by Chetrite and Mallick~\cite{Chetrite2012}.
Fluctuation theorems under unital CPTP maps for thermodynamic quantities, like work, energy and information-theoretic entropy, have appeared in numerous works~\cite{Rastegin2013,Rastegin:2014hc,Albash:2013fq}, while predictions for non-unital CPTP maps usually take the form of an integral fluctuation theorem with a so-called correction~\cite{Albash:2013fq,Rastegin:2014hc,Kafri2012,Goold:2014vx,Goold2015}.

In this paper we present a general formalism based on a generalized detailed balance condition that includes extends many of the previous results without the need for a correction term. As a consequence, our result clarifies  the minimal hypotheses
 needed to derive a fluctuation theorem for quantum maps. Our theorem is independent of the physical nature of the process that induces the CPTP map. This is a relevant feature. It makes the fluctuation theorem general enough to be applied to situations far from equilibrium, like systems in contact with coherent reservoirs~\cite{Scully2003,Lutz2009,Horowitz:2013jd}. Moreover, such a general result could be useful to analyze the thermodynamics of quantum processes whose physical details are not completely known, such as decoherence or quantum collapse. 

The paper is organized as follows. In Sec.~\ref{sec:kraus} we review the theory of CPTP maps and the Kraus representation, introducing the dual map, necessary to state the fluctuation theorem. In Sec.~\ref{sec:FT}   Êwe prove the general theorem for single maps and for a series of concatenated maps. Some applications are discussed in Sec.~\ref{sec:apps}. Finally, in Sec.~\ref{sec:conc} we summarize our results and present the main conclusions of the paper.

\section{Quantum operations and dual dynamics}
\label{sec:kraus}

Consider a generic CPTP quantum map $\rho \rightarrow \rho'\equiv \E(\rho)$ acting on the density matrix $\rho$ of a quantum system.
Any CPTP map admits a Kraus representation in terms of a collection of linear operators $\{M_k\}$ as \cite{kraus,breuer}
\begin{equation} \label{map}
 \E(\rho) = \sum_k \E_k(\rho)=\sum_k M_k \rho M_k^\dagger,
\end{equation}
with  $\sum_k M_k^\dagger M_k = \mathbb{1}$, a condition that ensures the trace-preserving property of the map $\E$.  
It is important to stress that the choice of $\{M_k\}$ is not unique: any family of operators  $M'_l = \sum_k U_{l k} M_k$, with $U_{l k}$  the entries of a unitary matrix, is also a valid Kraus representation. Not even the number of Kraus operators is unique for a given map.
For instance, if the Hilbert space of the system has finite dimension $N$, there exists a Kraus representation for any map with at most $N^2$ operators. However, using more than $N^2$ operators is sometimes necessary for a complete description of the physical process associated to the map (as we will see below).

\subsection{Quantum trajectories and nonselective states}

The Kraus representation (\ref{map}) is not just a mathematical way of writing the map; it also provides a physical picture of the map as a random transformation of pure states. A specific representation decomposes the map into a number of operations $\E_k(\cdot)=M_k\cdot M_k^\dag$. Each operation  transforms a pure state $\ket{\psi}$ into a new pure state 
\begin{equation}
\ket{\psi_k'}=\frac{M_k\ket{\psi}}{||M_k\ket{\psi}||},
\end{equation}
with probability $p_k(\ket{\psi})\equiv||M_k\ket{\psi}||^2$  ($\sum_k p_k(\ket{\psi})=1$). 
This picture extends to mixed states of the form $\rho=\sum_i  p_i \ket{\psi_i}\bra{\psi_i}$, which represents  a classical  ensemble of pure states  $\ket{\psi_i}$ each sampled with probability $p_i$.
Thus, the probability that operation $k$ occurs is
\begin{equation}
\begin{split}
p_k(\rho) &= \sum_i p_ip_k(\ket{\psi_i})= {\rm Tr}[\E_k(\rho)]
\end{split}
\end{equation}
and the final state conditioned on this operation is
\begin{equation}
\rho'_k=\sum_i  p_i \frac{M_k\ket{\psi_i}   \bra{\psi_i} M_k^\dagger}{||M_k\ket{\psi_i}||^2} =\frac{\E_k(\rho) }{p_k(\rho)}.
\end{equation}
If we know which operation $\E_k$ has occurred, then $k$ can be seen as the outcome of a generalized measurement and 
$\rho'_k$ as the \emph{selective post-measurement state} of the system. If we do not know which operation took place (or we decide not to incorporate that information into our description), then the state after the transformation is  $\rho'=\E(\rho)=\sum_kp_k(\rho)\rho'_k$, 
 usually referred to as the {\em nonselective post-measurement state}   Ê(although the transformation given by the map $\rho'=\E(\rho)$ does not necessarily
  imply  any measurement and not even  a specific Kraus representation). 
This setup defines an {\em efficient generalized measurement} in quantum mechanics, more restrictive than generalized measurements where the observer has access only to a function $f(k)$ of the operation index $k$, which may not be one-to-one \cite{wiseman}. 

A generic quantum evolution is described by a concatenation of maps $\E_r$ with Kraus operators $M^{(r)}_k$. For the initial state $\rho(0)$, the nonselective state evolves as
\begin{equation}
\rho(r)=\E_r\E_{r-1}\dots \E_1\rho(0).
\end{equation}
This density matrix $\rho(r)$ can be interpreted as the average of the stochastic evolution. 
If the initial state is pure $\rho(0)=|\psi(0)\rangle\langle\psi(0)|$, then a stochastic trajectory $\gamma\equiv (k_1,k_2,\dots k_r)$ is given by the operations $k_r$ that occurred in the application of map $\E_r$ and determines the evolution of the pure state:
\begin{equation}
\ket{\psi(r)}= M^{(r)}_{k_r}M^{(r-1)}_{k_{r-1}}\dots M^{(1)}_{k_1}\ket{\psi(0)}.
\end{equation}

\subsection{Dual dynamics}

Now consider a particular Kraus representation of a map $\E=\sum_k\E_k$, and suppose that the map has a positive-definite invariant state $\pi$ (not necessarily unique), i.e., $\E(\pi) = \pi$. 
For such maps, we introduce an auxiliary or {\it dual} map ${\tilde \E}$ with respect to $\pi$ and to a fixed, arbitrary unitary or anti-unitary operator ${\A}$.
Inspired by Crooks, we define this dual map through the equality \cite{Crooks:2008ji,Horowitz:2013jd}
\begin{equation}\label{dual0}
{\rm Tr}\left[\E_{k_2}\E_{k_1}(\pi)\right]={\rm Tr}\left[\tilde\E_{k_1}\tilde\E_{k_2}(\tilde\pi)\right]
\end{equation}
where $\tilde\pi\equiv \A \pi \A^\dagger$ is the invariant state transformed by $\A$. Equation~\eqref{dual0}  states that  the probability of observing the outcome $k_1$ followed by $k_2$ when we apply the map twice to the invariant state $\pi$ equals the probability of observing the reverse outcome ---$k_2$ followed by $k_1$--- when the dual map is applied twice to $\tilde\pi$. 
In this way, the dual map induces a dynamics in the invariant state that is the  reverse of the original one. 
Following the derivation introduced by Crooks in Ref.~\cite{Crooks:2008ji}, one can prove  Êthat the Kraus operators of the dual map are given by
\begin{equation}\label{Kraus_dual}
\tilde {M}_k  \equiv \A\pi^{\frac{1}{2}} M_k^\dagger \pi^{-\frac{1}{2}}  \A^\dagger.
\end{equation}
 Trace preservation ($\sum_k \tilde{M}^{{\dagger}}_k {\tilde M}_k = \mathbb{1}$) follows immediately from $\E(\pi) = \pi$, and one can verify that the dual map preserves the dual invariant state, $\tilde\E(\tilde\pi)=\tilde\pi$.

The inclusion of the operator ${\cal A}$ in the definition of the dual map is not mathematically necessary to derive the fluctuation theorem. In fact, ${\cal A}$ does not appear in the original definition by Crooks \cite{Crooks:2008ji}. However, in some situations an appropriate choice of the operator ${\cal A}$ is needed to find a dual dynamics with a precise physical interpretation
 or that is suitable of being implemented in the laboratory.
The customary choice  
is  the time-reversal operator $\A=\Theta$ that changes the sign of odd variables, like linear and angular momenta. $\Theta$ is an anti-linear, anti-unitary operator, satisfying $\Theta^2=\Theta^\dagger\Theta=\Theta\Theta^\dagger=\mathbb{1}$~\cite{Haake,Campisi:2011ka}. For instance, $\Theta$  acts on a spinless particle by complex conjugation of the wave function in the position representation. The need of $\Theta$ in the definition of the dual process is clear, for example, if the map is a unitary evolution, i.e., a map given by a unique Kraus operator  ${ U}$ with ${ U}^\dagger ={ U}^{-1}$. In that case the invariant state is proportional to the identity matrix and the dual dynamics reads
\begin{equation}
\tilde { U} = \Theta \, { U}^{\dagger} \Theta^{\dagger}.
\end{equation}
The dual map is again a unitary evolution given by the unitary operator $\tilde {U}$ and corresponds to the {\em operational} time reversal of the original unitary evolution given by ${ U}$~\cite{Andrieux:2008em}. For instance, if ${U}$ is the evolution of a system under a constant Hamiltonian $H$, ${ U}=e^{-{\rm i}Ht/\hbar}$, and $H$ is time-reversal invariant, $[ H,\Theta]=0$, then $\tilde{ U}={ U}$, i.e., the dual map is identical to the original one. On the other hand, if the Hamiltonian depends on time according to some protocol, and ${ U}$ is the evolution between $t=0$ and $t=\tau$, then $\tilde { U}$ is the evolution that results when the protocol is reversed (which is, in general, different from $U^{\dagger}$).

The operator $\A$ can also account for other transformations of the system state that are necessary to exploit dynamical and static symmetries. 
In fact, this freedom  has a classical counterpart in fluctuation theorems that incorporate various symmetry transformations~
\cite{Maes:1999kn,Hurtado:2011gm,Lacoste:2014kh}.

 \section{Fluctuation theorems}\label{sec:FT}

\subsection{Nonequilibrium potential and detailed balance}

We now prove a general fluctuation theorem for a large family of CPTP maps.
To begin our introduction of these maps, let us focus on an important class of maps that admit the following Kraus representation
\begin{equation}\label{classical}
 M_{ji} = \alpha_{ji} \ket{\pi_j} \bra{\pi_i},
\end{equation}
in terms of the eigenstates $\{\ket {\pi_i}\}$ of the invariant density $\pi$.
Here the Kraus operators are labeled by two indices $(i,j)$ that identify  jumps or transitions between eigenstates of $\pi$, $\ket{\pi_i}\to \ket{\pi_j}$, occurring with probability $||M_{ji}\ket{\pi_i}||^2=|\alpha_{ji}|^2$. 
These maps are special in that a single application of $\E$  destroys any coherences between eigenstates of $\pi$ in the initial state $\rho$, reducing the subsequent action of the map to a classical Markov chain on the eigenstates $\{\ket{\pi_i}\}$. 
Therefore, the dynamics induced by CPTP maps of the form \eqref{classical} Êis essentially classical. On the other hand, quantum effects arise if the Kraus operators are linear combinations of the transition operators $\ket{\pi_j} \bra{\pi_i}$, preserving coherences between eigenstates of the invariant density matrix.

The family of maps that obey a fluctuation theorem go slightly beyond the ``classical'' case outlined above \eqref{classical}. 
To make this family precise, we assign to each eigenstate $\ket{\pi_i}$, whose strictly positive eigenvalue is denoted by $\pi(i)$, a {\em nonequillibrium  potential}, similar to the one used in the classical Hatano-Sasa theorem \cite{Hatano:2001uc},
\begin{equation}\label{potentialdef}
\Phi_\pi(i) \equiv -\ln \pi(i).
\end{equation}
Then the maps that obey our fluctuation theorem are those where each Kraus operator $M_k$ is formed from a superposition of  jump operators, all of them inducing the same change in nonequilibrium potential  $\Delta\Phi_\pi(k)$:
\begin{equation}\label{Kraus_rep}
 M_k = \sum_{i, j} m_{j i}^k \ket{\pi_j} \bra{\pi_i} ,
\end{equation}
with $m_{j i}^k = 0$ if $\Phi_\pi(j) - \Phi_\pi(i) \neq \Delta \Phi_\pi(k)$. That is, by measuring the operation $M_k$ we know without uncertainty the change in the nonequilibrium potential, even though that change could have occurred through a superposition of jumps. 
One example of this construction is a harmonic oscillator coupled to an equilibrium reservoir of resonant photons at temperature $T$ \cite{Horowitz:2012gn}. 
Here, the nonequilibrium potential is the energy of each eigenstate, divided by $kT$, and the change in the nonequilibrium potential in a transition is  proportional to the energy transferred to the reservoir of photons as heat. 
By measuring the reservoir we are able to detect jumps in the oscillator, but the measurement, in general, does not provide  information about the system state \cite{Horowitz:2012gn}.

It is straightforward to check that condition (\ref{Kraus_rep}) is equivalent to
\begin{eqnarray}
[ M_k,\ln\pi ] &=& \Delta\Phi_\pi(k) M_k \nonumber \\
{[} M^\dagger_k,\ln\pi ] &=& -\Delta\Phi_\pi(k)M^\dagger_k \label{comm}
\end{eqnarray}
and, consequently $[M^\dagger_kM_k,\ln\pi]=[M^\dagger_kM_k,\pi]=0$. These commutation relations are similar to those satisfied by the Lindblad operators that appear in Davies' theory of systems weakly coupled to thermal baths (see below and \cite{Alicki:2006tr,Szczygielski:2013wc,Rivas:2012wd}). 
They indicate that the pair $M_k$, $M^\dagger_k$ acts as ladder operators, inducing jumps between the eigenstates $\ket{\pi_i}$ of $\pi$ with a fixed change $\Delta\Phi_\pi(k)$ in the nonequilibrium potential $\Phi$.
Finally, (\ref{Kraus_rep}) ensures that the dual Kraus operators obey a generalized detailed balance condition
\begin{equation} \label{dual_Kraus}
 {\tilde M}_{k} = e^{\Delta \Phi_{\pi}(k)/2}\, \A M_{k}^\dagger  \A^\dagger
\end{equation} 
that can be obtained by plugging  (\ref{Kraus_rep}) into (\ref{Kraus_dual}). One can also prove that the form \eqref{Kraus_rep} is the only one for which the dual operators $\tilde M_k$ in (\ref{Kraus_dual}) are proportional to $\A M_{k}^\dagger  \A^\dagger$. Remarkably, for maps with multiple invariant states the $\Delta \Phi_\pi(k)$ do not depend on the specific invariant state $\pi$ chosen to define the nonequilibrium potential and the dual dynamics~\footnote{F. Fagnola, private communication}. In other words, the set of values $\Delta \Phi_\pi(k)$ is a property of the map $\E$.

\subsection{Fluctuation theorem for a single CPTP map}

The basis of our fluctuation theorem is codified in the proportionally between Kraus operators and their dual counterpart in (\ref{dual_Kraus}).
This generalized detailed balance condition connects the probability to observe a given jump, say $k$, with the probability to observe the same jump in the dual dynamics.
Specifically, suppose that we initially prepare the system in the pure state $\ket{\psi_n}$, and then apply the map $\E$, registering the occurrence of the operation $k$.
We then perform a quantum yes/no measurement of the projector $\ket{\phi_m}\bra{\phi_m}$.
The subscripts $n$ and $m$ are added to the initial and final states so that later on we can consider measurements of arbitrary observables with eigenstates $\ket{\psi_n}$ and $\ket{\phi_m}$.

Now, let $p (m,k|n)$ be the probability that given an initial state $\ket{\psi_n}$ we observe operation $k$ {\em and} the final state $ \ket{\phi_m}$, that is, the probability to observe the jump $\ket{\psi_n} \rightarrow \ket{\phi_m}$ under the action of $M_k$. Let $\tilde{p} (n, k | m)$ be the probability to observe the inverse jump  $\ket{\tilde{\phi}_m} \rightarrow \ket{\tilde{\psi}_n}$, with $\ket{\tilde{\psi}} = \A \ket{\psi}$, under the action of the dual operator $\tilde M_k$. 
Using (\ref{dual_Kraus}), the ratio of these two conditional probabilities is
\begin{eqnarray} \label{micro-reversibility}
 \frac{p (m, k | n)}{\tilde{p} (n, k | m)} &=& \frac{|\bra{\phi_m} M_k \ket{\psi_n}|^2}{|\bra{\tilde{\psi}_n} \tilde{M}_k  \ket{\tilde{\phi}_m}|^2} 
 =\frac{|\bra{\phi_m} M_k \ket{\psi_n}|^2}{|\bra{{\psi}_n}\A^\dagger \tilde{M}_k\,  \A\ket{{\phi}_m}|^2}\nonumber\\ 
&=& \frac{|\bra{\phi_m} M_k \ket{\psi_n}|^2}{|\bra{\psi_n} M_k^\dagger \ket{\phi_m}|^2}\, \frac{1}{e^{\Delta \Phi_\pi (k)}} 
 = e^{-\Delta \Phi_\pi(k)} 
\end{eqnarray}
 Equation (\ref{micro-reversibility}) can be considered as a modified detailed balance relation for 
the operation $\E_k$ and its dual $\tilde{\E}_k$, which remarkably is independent of the initial and final states. 

Suppose now that we prepare the system in the initial mixture $\rho_{\rm i} = \sum_n p_{\rm i}(n) \ket{\psi_n} \bra{\psi_n}$ and apply the map $\E$. 
By measuring the initial state $\ket{\psi_n}$, the operation $\E_k$ and a final state $\ket{\phi_m}$ we obtain a {\em trajectory} $(m,k,n)$ that is observed with a probability $p(m,k,n)=p(m,k|n)p_{\rm i}(n)$. 
We compare this to a dual process induced by the map $\tilde\E$ applied to the initial state $\tilde{\rho}_{\rm f} = \sum_m \tilde p_{\rm f} (m) \ket{\tilde{\phi}_m} \bra{\tilde{\phi}_m}$. 
The dual trajectory $(n,k,m)$ is given as well by the initial state $\ket{\tilde\phi_m}$, the dual operation $\tilde \E_k$ and the final state $\ket{\tilde\psi_n}$, and it is observed with probability $\tilde p(n,k,m)=\tilde p(n,k|m)\tilde p_{\rm f}(m)$.  The ratio of the probability to observe a trajectory $(n,k,m)$ and the probability to observe the reverse trajectory $(m,k,n)$ in the dual process is then, from (\ref{micro-reversibility}),
\begin{eqnarray}\label{detailed_ft}
\Sigma(n,k,m)\equiv\ln{\frac{p(n,k,m)}{\tilde{p}(m, k ,n)}} =  \sigma(n,m) - \Delta \Phi_{\pi}(k),
\end{eqnarray}
where $\sigma(n, m) \equiv -\ln{\tilde p_{\rm f}(m) + \ln{p_{\rm i}(n)}}$ is a boundary term, only depending on the initial state of the process $\rho_{\rm i}$ and the initial state of the dual $\tilde \rho_{\rm f}$. The quantity $\Sigma$ is a measure of how different the original and the dual trajectories are. In particular, when the dual is the time reversed process (see below), $\Sigma$ is a measure of the irreversibility of the process for a given trajectory. In the rest of the paper we will show that it can be identified with an entropy production in many situations of  interest.

 A Jarzynski-type intergral fluctuation theorem  
 immediately follows from  \eqref{detailed_ft}:
\begin{equation} \label{integral_ft}
\left\langle e^{-\Sigma} \right\rangle = 1,
 \end{equation}
 where the average is over forward trajectories, $p(n,k,m)$.
Finally by Jensen's inequality $\langle e^x \rangle \geq e^{\langle x \rangle}$, we have the second-law-like inequality 
\begin{equation}\label{positive}
\langle\Sigma\rangle=\braket{\sigma} -   \braket{\Delta \Phi_\pi} \geq 0.
\end{equation}

\subsection{Fluctuation theorem for concatenated maps}

Our fluctuation theorems (\ref{detailed_ft}--\ref{integral_ft}) can be easily extended to a concatenation of CPTP maps, $\Omega=\E_R\E_{R-1}\dots\E_r\dots\E_1$, which is the case of general Markov quantum evolution, unitary evolution punctuated by projective measurements, driven systems in contact with thermal baths, etc. A trajectory now is given by the initial $\ket{\psi_n}$ and final states $\ket{\phi_m}$ and the outcomes $k_r$ of all the measurements associated to the maps $r=1,2,\dots,R$:  $\gamma=(n,k_1,k_2,\dots,k_R,m)$. Each map $\E_r$ has a Kraus representation, given by the operators $M_k^{(r)}$, and an  invariant state $\pi^{(r)}$ for which the dual map $\tilde \E_r$ and the nonequilibrium potential $\Phi_{\pi^{(r)}}(i)$ are defined as in Eqs.~\eqref{Kraus_dual} and \eqref{potentialdef}.

To derive the fluctuation theorem, we reverse the concatenation of maps. We define  the dual process as  $ \hat \Omega=\tilde \E_1\dots\tilde \E_r\dots\tilde \E_{R-1}\tilde \E_{R}$  (notice that, for $R>1$, in general, $\hat\Omega\neq \tilde \Omega$, i.e., the dual process does not coincide with the dual map of $\Omega$). 
If each map obeys condition \eqref{Kraus_rep} [or, equivalently, \eqref{dual_Kraus}], then we get the following symmetry relation
\begin{eqnarray}
\frac{p(m,k_R,\dots,k_1 |n)}{\tilde{p}(n, k_1,\dots,k_R | m) } &=& \frac{|\bra{\phi_m} M_{k_R}^{{(R)}} \dots 
M^{(1)}_{k_1} \ket{\psi_n}|^2}{|\bra{\tilde\psi_n}  \tilde{M}^{(1)}_{k_1}\dots  \tilde{M}^{(R)}_{k_R}  \ket{\tilde\phi_m}|^2}\nonumber \\
&=& \exp\left[-\sum_{r=1}^R \Delta\Phi_{\pi^{(r)}}(k_r)\right]
\end{eqnarray}
A detailed fluctuation theorem can be now obtained by comparing the probability of a trajectory $\gamma=(n,k_1,\dots,k_R,m)$ in the forward process  and  the probability of the inverse trajectory
$\tilde{\gamma}=(m,k_R,\dots,k_1,n)$ in the dual process:
\begin{equation}\label{detailed_ft_concatenation}
\Sigma(\gamma) \equiv \ln{\frac{p(\gamma)}{\tilde p(\tilde{\gamma})}} =  \sigma(n,m) - \sum_{r=1}^R \Delta \Phi_{\pi^{(r)}}(k_r),
\end{equation}
with a corresponding integral fluctuation theorem that follows readily, like in \eqref{integral_ft}.
Thus, for a concatenation of maps implemented in sequence, we merely have to add the changes in the nonequilibrium potential along the trajectory.
Notice also that we effectively used a Kraus representation for the map $\Omega$ where each Kraus operator was labeled with the sequence $\{k_1,\dots,k_R\}$, requiring possibly many more than the necessary $N^2$ operators.

A clear interpretation of $\Sigma(\gamma)$ arises if we consider the concatenation of the same map $\E$, acting on the stationary density matrix $\pi$, and the corresponding dual process acting on $\tilde \pi$. In this case $p_{\rm i}(n)=\pi_n$ and $\tilde p_{\rm f}(m)=\pi_m$, yielding
\begin{equation}
\Sigma(\gamma) = \ln \pi_n -\ln \pi_m -  \sum_{r=1}^R \Delta \Phi_{\pi}(k_r)=0
\end{equation}
for any trajectory $\gamma$. This is expected from the (modified) Crooks definition \eqref{dual0}: the original and the dual maps acting on $\pi$ and $\tilde\pi$, respectively, produce a trajectory $\gamma$ and its reverse $\tilde\gamma$ with identical probability. Therefore, $\Sigma$ can be considered as a measure of the distinguishability of the original and the dual process, but also as a measure of how far the system is from the stationary state. These two equivalent interpretations are familiar in thermodynamics when $\pi$ is an equilibrium state: the dual is the reverse process and $\Sigma(\gamma)$ is the entropy production which measures both irreversibility and departure from equilibrium~\cite{Kawai:2007kc}. In more general situations, $\Sigma(\gamma)$ is the part of the entropy production due to the fact that the state of the system does not coincide with the stationary state. This can occur in the transient from a nonsteady initial condition to the stationary state, or due to a finite-speed, or nonadiabatic, driving. In any case, $\Sigma(\gamma)$ is known as the nonadiabatic  \cite{Esposito:2010jf,Esposito:2010vu,Esposito:2010to} or  excess   \cite{Hatano:2001uc,Chernyak2006b} entropy production, in contrast with the entropy production needed to maintain the stationary state, which is often referred to as adiabatic or house-keeping entropy production~\cite{Speck2005b}.

The fluctuation theorem stated in \eqref{detailed_ft_concatenation} exploits the dynamical symmetries of the process through the dual map and the nonequilibrium potential, in the same spirit as the detailed fluctuation theorem for processes connecting nonequilibrium states developed by Esposito and Van den Broeck \cite{Esposito:2010jf,Esposito:2010vu,Esposito:2010to}. Finally, the integral theorem \eqref{integral_ft} is the quantum version of the Hatano-Sasa theorem \cite{Hatano:2001uc}, extending the Jarzynski equality to nonequilibrium states. 
The corresponding second-law-like inequality~\eqref{positive} extends to arbitrary boundary conditions the quantum Hatano-Sasa inequality for concatenated CPTP maps proposed by Sagawa~\cite{Sagawa:2012vd}.

\section{Applications}\label{sec:apps}

Despite their simplicity, the above fluctuation theorems include as special cases many of the known quantum fluctuation relations.
In the section, we explain how these relations come about in our formalism. We first discuss the boundary term $\sigma(n,m)$ and then apply the general theorem to different dynamics. Here we specify $\A=\Theta$, the anti-unitary time-reversal operator.

\subsection{Boundary terms}

There are two common choices for boundary terms: {\em i)}  setting the initial state of the dual equal to the final state of the forward process $\tilde\rho_{\rm f}=\A \rho_{\rm f}\A^\dagger$, $\rho_{\rm i}$ being an arbitrary state; and {\em ii)}
setting the initial state $\rho_{\rm i}$ of the forward process  and the initial state $\tilde\rho_{\rm f}$ of the dual process as equilibrium states. 
Notice that by selecting the initial states of the forward and dual processes we are also fixing the basis in which the quantum measurements are performed at the beginning and end of the processes.

In the first case, the boundary term
\begin{equation}\label{entropy_sigma}
\sigma(n,m)=-\ln{ p_{\rm f}(m) + \ln{p_{\rm i}(n)}}=s_{\rm f}(m)-s_{\rm i}(n)
\end{equation}
is the increase of the stochastic or trajectory entropy \cite{Seifert:2005fu,Horowitz:2012gn,Horowitz:2013jd,Monnai2005}, whose average over forward trajectories yields the increase of von Neumann entropy. 

This choice is relevant from a theoretical point of view, but the resulting dual process is hard to implement in general, except when 
the system is small enough to be prepared in an arbitrary state (say, a few qubits or a harmonic oscillator).

The second choice, equilibrium initial states for the forward and dual dynamics, is more interesting from  an operational  point of view, since the dual dynamics can be easily implemented in the laboratory by equilibrating the system with a thermal reservoir and reversing the protocol that drives the Hamiltonian \cite{Campisi:2011ka,Albash:2013fq,Batalhao:2014ta}. Let us suppose that, before applying any quantum map, the system Hamiltonian is initially fixed $H_{\rm i}$, whereas after the Hamiltonian is $H_{\rm f}$.
We further take the initial state of the forward process to be equilibrium at inverse temperature $\beta$, that is, $\rho_{\rm i}=e^{\beta(F_{\rm i}- H_{\rm i})}$, where $F_{\rm i}$ is the corresponding free energy.
Similarly, we initialize the dual process in the final equilibrium at the same temperature, $\tilde\rho_{\rm f}=e^{\beta(F_{\rm f}- H_{\rm f})}$.
Then,
\begin{equation}\label{work_sigma}
\sigma(n,m)=\beta (E^{\rm f}_m-E^{\rm i}_n-F_{\rm f}+F_{\rm i}) \equiv\beta( \Delta E_{n,m}-\Delta F)
\end{equation}
where the $\{E^{\rm i,f}_l\}$ are the eigenvalues of the initial and final Hamiltonians, respectively.

\subsection{Unital work relations} 

As a first example, we take our quantum map to be \emph{unital} (or bistochastic~\cite{Bengtsson}), that is, the identity is an invariant state, $\E(\mathbb{1})=\mathbb{1}$  (although the identity may not be the only one).  Any unitary evolution $U$ is unital, $U \mathbb{1} U^\dagger=\mathbb{1}$, and its dual map is the time-reversal $ \tilde U = \Theta U^\dagger \Theta^\dag$. 
Another example of a unital map is the projective measurement of an observable but, more generally, any minimally disturbing measurement is unital \cite{wiseman}.
For these maps, the Kraus operators are self-adjoint $M_k^\dagger =M_k$, leading to dual operators $\tilde M_k=\Theta M_k\Theta^\dagger$.
Finally, pure decoherence is also implemented with unital maps that remove all the off-diagonal elements in a specified basis.
For all such unital maps or concatenation of such maps, $\Delta \Phi_\pi(k)=0$ for all $k$, and the fluctuation theorem only consists of the boundary term.

Let us now consider a concatenation of unital maps as describing a physical process. An important example is a process consisting of  several unitary transformations induced by driven time-dependent Hamiltonians, punctuated by a number of measurements and/or pure decoherence processes. In each map, energy can be transferred to the system. We call the energy input into the system due to the driving $W_{\rm drive}$, driving work, and $W_{\rm meas}$ the energy input due to the measurements and/or decoherence processes. Whereas the driving work $W_{\rm drive}$ has a clear interpretation as the energy supplied by driving, the origin of the energy input due to measurement is still obscure. This energy transfer occurs, for instance, in a projective measurement of an observable that does not commute with the Hamiltonian. In any case,  $\Delta E_{n,m}= W_{\rm drive} + W_{\rm meas}$ and, if we choose equilibrium initial states the boundary term $\sigma$ is given by \eqref{work_sigma} 
and
\begin{equation}\label{unitalft}
\Sigma(\gamma) =  \beta(W_{\rm drive}+W_{\rm meas}-\Delta F)=\beta W_{\rm diss}.
\end{equation}
The fluctuation theorem \eqref{detailed_ft_concatenation}, therefore, 
reproduces the work fluctuation theorems for unital processes derived in \cite{Campisi:2010kx, Watanabe:2014fh,Rastegin2013,Albash:2013fq} (see also~\cite{Callens2004, DeRoeck2004}). Notice that, if we allow the system to relax to equilibrium after the maps have been applied, then $\Sigma(\gamma)$ equals the entropy production along the whole process. We stress that this result is valid for any concatenation of unital maps.
On the other hand, if we choose the initial state of the dual process as the final state of the original process, $\Sigma = -\ln p_{\rm f}(m)/p_{\rm i}(n)$ is just the change in stochastic entropy.
When averaged, the entropy production $\Sigma$ becomes the change in the von Neumann entropy of the system
\begin{equation}
\Delta S_{\rm sys}=\langle \Sigma\rangle\ge 0,
\end{equation}
whose positivity follows from \eqref{positive}.  This provides another proof of the well-known property that unital maps can only increase the von Neumann entropy~\cite{breuer}.

\subsection{Thermalization and heat} 

Another interesting example is a generic thermalization map~\cite{Crooks:2008ji} at inverse temperature $\beta=1/(kT)$ (or Gibbs-Preserving map~\cite{Faist2015}), that is, a map  whose  invariant state is the equilibrium density matrix  $\pi=e^{\beta(F- H)}$, where $H=\sum_j E_j\ket{e_j}\bra{e_j}$ is the Hamiltonian of the system and $F$ its free energy at temperature $T$.
Thus, the nonequilibrium potential is related to the energy as $\Phi_\pi(j)=-\ln\pi(j)=\beta (F- E_j)$.
To verify our fluctuation theorem, each Kraus operator $M_k$ must promote transitions between energy eigenstates involving a given change of energy $\Delta E(k)$, that is, $M_{k}=\sum_{ji}m^k_{ji}\ket{e_j}\bra{e_i}$, where the sum runs over pairs of energy eigenstates with the same energy difference $\Delta E(k)=E_j-E_i$.
Now, since the energy is supplied by a thermal reservoir, we can identify these energy exchanges as heat flowing to the reservoir, $Q(k)=-\Delta E(k)$.
The dual Kraus operators ${\tilde M}_k \propto M_k^\dag=\sum_{ji}m^k_{ji}\ket{e_i}\bra{e_j}$ (for a time-reversal invariant $H$) induce the reverse transitions accompanied by the reverse flow of heat ${\tilde Q}(k)=-Q(k)$, and thus can be identified with a Kraus operator in the original map.

We now can consider a thermodynamic process formed by a concatenation of thermalization steps by $N$ distinct thermal reservoirs with inverse temperatures $\{\beta_i\}_{i=1}^N$ interspersed by unital transformations (unitary drivings, measurements or decoherence). 
For this setup, if we choose the initial state of the dual process as the final state of the original process, we arrive at 
\begin{equation}
\Sigma(\gamma)=s_{\rm f}(m)-s_{\rm i}(n)+\sum_{i=1}^N\beta_i Q_i(\gamma),
\end{equation}
with $Q_i(\gamma)$ the total heat flow into the $i$-th reservoir. In this case, we get a fluctuation theorem for the total irreversible entropy production in the process.

On the other hand, the equilibrium boundary terms are interesting when restricted to one thermal reservoir, leading to 
\begin{equation} \label{Crooks-Jarzynski_FT}
\Sigma(\gamma)=\beta(\Delta E_{n,m}-\Delta F+Q(\gamma)) = \beta(W(\gamma)-\Delta F),
\end{equation}
employing the energy balance $\Delta E_{n,m}=W(\gamma)-Q(\gamma)$. Again, $\Sigma$ equals the entropy production along the whole process consisting of the map concatenation followed by a thermal relaxation. 
 The detailed and integral fluctuation theorems following from the identification \eqref{Crooks-Jarzynski_FT} are respectively the quantum Tasaki-Crooks and Jarzynski fluctuation theorems for thermal maps punctuated by unital maps~\cite{Campisi:2011ka}.

\subsection{Lindblad master equations}

Another nice illustration of our results are the Lindblad master equations that model the Markovian dynamic evolution of open quantum systems \cite{breuer,Rivas:2012wd}.
For a quantum system with Hamiltonian $H$, a Lindblad master equation is specified by a collection of positive Lindblad operators $\{L_k\}_{k=1}^K$ as
\begin{equation}\label{Lindblad}
\partial_t \rho_t = -i[H,\rho_t] + \sum_k{\mathcal D}[L_k]\rho_t\equiv {\mathcal L}\rho_t,
\end{equation}
where the dissipator ${\mathcal D}$ is defined as ${\mathcal D}[L]\rho = L\rho L^\dag-\frac{1}{2}\left(L^\dag L\rho+\rho L^\dag L\right)$.
To make contact with our fluctuation theorem, we begin by observing that the solution to \eqref{Lindblad} can be obtained by concatenating a sequence of maps together that evolve the system forward in small time steps $d t$:
\begin{equation}\label{eq:dtLindblad}
\begin{split}
\E(\rho_t) &= (\mathbb{1}+{\mathcal L}d t)\rho_t =  M_0\rho_t M_0^\dag +\sum_{k=1}^K M_k\rho_t M_k^\dag,
\end{split}
\end{equation}
with Kraus operators
\begin{eqnarray}
M_0 &= &\mathbb{1} -\Big(iH+\frac{1}{2}\sum_k L_k^\dag L_k\Big)d t  \\
M_k &= & L_k \sqrt{d t}, \qquad 1\le k \le K.
\end{eqnarray}
This map has at least one invariant state $\pi$ \cite{Rivas:2012wd}, obeying ${\mathcal L}\pi=0$.

To satisfy our fluctuation theorem, the Kraus operators $\{M_k\}$ must be of the form \eqref{Kraus_rep} and verify the generalized detailed balance relations \eqref{dual_Kraus}.
Enforcing these conditions on $\{M_k\}_{k\ge 1}$ immediately leads to a restriction on the Lindblad operators similar to \eqref{Kraus_rep}.
Namely, each Lindblad operator must induces jumps between invariant-state eigenstates, $L_k=\sum_{ji}m^k_{ji}\ket{\pi_j}\bra{\pi_i}$, where $m^k_{ji}=0$ for all $i,j$ such that $\Phi_\pi(j)-\Phi_\pi (i) \neq \Delta\Phi_\pi(k)$. In this case, the generalized detailed balance relation \eqref{dual_Kraus} holds:
\begin{equation} \label{dual_Lindblad}
 {\tilde L}_{k} = e^{\Delta \Phi_{\pi}(k)/2}\, \Theta L_{k}^\dagger  \Theta^\dagger, \qquad k\ge 1.
\end{equation}
As for the Kraus operators, if the Lindblad operator $L_k$ induces jumps where the nonequilibrium potential change equals $\Delta\Phi_\pi(k)$, then they obey commutation relations similar to \eqref{comm}:
\begin{eqnarray}\label{comml}
[ L_k,\ln\pi ] &=& \Delta\Phi_\pi(k) L_k \nonumber \\
 {[}L^\dagger_k,\ln\pi ] &=& -\Delta\Phi_\pi(k)L^\dagger_k,
\end{eqnarray}
and $[L^\dagger_k L_k,\ln\pi]=[L^\dagger_k L_k,\pi]=0$.

Let us verify now whether $M_0$ also satisfies our conditions. The dual operator \eqref{Kraus_dual} reads:
\begin{align}
{\tilde M_0} = \Theta \pi^{\frac{1}{2}} \Big[\mathbb{1} -\Big(-iH+\frac{1}{2}\sum_k L_k^\dag L_k\Big)d t\Big] \pi^{-\frac{1}{2}}\Theta^\dag
\end{align}
Since $[L^\dagger_k L_k,\pi]=0$, for our generalized detailed balance condition to hold, that is ${\tilde M}_0\propto \Theta M_k^\dag \Theta^\dag$, we must assume that $[H,\pi]=0$, forcing the invariant state to be diagonal in the energy eigenbasis.
An immediate consequence of this observation is that in fact the $\Delta \Phi_\pi$ must correspond to jumps in the energy.
With this additional assumption, we have 
\begin{eqnarray}\nonumber
{\tilde M}_0 &=& \Theta\Big[\mathbb{1} -\Big(-iH+\frac{1}{2}\sum_k L_k^\dag L_k\Big)d t\Big]\Theta^\dag \\  &=& \Theta M_0^\dag \Theta^\dag 
\end{eqnarray}
Thus, $M_0$ satisfies our generalized detailed balance relations with $\Delta \Phi_\pi(0)=0$, as one would expect for a Kraus operator that does not induce transitions.
The restrictions on the Lindblad operators outlined here as assumptions can in general be proved as consequences of the requirement that the dual map also be CPTP~\cite{Fagnola:2007hj}.

Consider now the following process. We run the Lindbladian evolution for an interval of time $[0,\tau]$, and measure some observables at time $t=0$ and $t=\tau$. In this scenario, a trajectory $\gamma=(n,k_1,k_2,\dots,k_N,m)$ is given by the initial and final measurement outcomes, $n$ and $m$ respectively, and a set of jumps $k_l$ occurring at times $t_l$. Notice that the stochastic trajectory, as defined in the previous sections, should contain a big number of instances $k_r=0$, i.e., corresponding to operation $M_0$, between jumps. However, these operations do not contribute to $\Sigma(\gamma)$ and we can omit them from the discussion. In this case,
\begin{equation}\label{eq:SigmaLindblad}
\Sigma(\gamma) = \sigma(n,m) +\sum_l \Delta\Phi_\pi (k_l),
\end{equation}
With the entropic boundary conditions~\eqref{entropy_sigma}, we arrive at the quantum generalization of the Hatano-Sasa theorem \cite{Hatano:2001uc} for the nonadiabatic entropy production of Lindblad master equations, as developed in~\cite{Horowitz:2013jd}. 
Its average over trajectories, corresponds to the integrated expression first introduced by Spohn for arbitrary quantum dynamical semigroups~\cite{Spohn1978}, then extended by Yukawa to driven quantum Markov processes~\cite{Yukawa:2001tf}.
The equivalence between our trajectory picture and the average thermodynamics behavior has been discussed in~\cite{Horowitz2014b}.

So far we have been treating the dissipation in the Lindblad master equation as a whole. 
When the dissipation can be interpreted as coming from $M$ distinct thermodynamic reservoirs (or Markovian noise processes), we can employ our formula for the entropy production of concatenated maps \eqref{detailed_ft_concatenation} to arrive at a complementary formulation of the thermodynamics.
The effect of each of the $M$ reservoirs is captured in the dynamics by a separate collection of Lindblad operators $\{L_{k,\alpha}\}_{k=1}^{K_\alpha}$, where $\alpha=1,\dots, M$ labels the reservoir:
\begin{equation}
\partial_t \rho_t = -i[H,\rho_t] + \sum_\alpha\sum_{k}{\mathcal D}[L_{k,\alpha}]\rho_t.
\end{equation}
Similar to before \eqref{eq:dtLindblad}, we can implement the evolution of this equation over a small time interval $dt$ by a map, except now it is formed by a \emph{concatenation} of intermediary maps, $\E(\rho_t)=\E_{\alpha_M}\cdots \E_{\alpha_1} \E_0(\rho_t)$, each arising from the different dynamical influences.
The first map $\E_0(\rho_t)= \rho_t - i[H, \rho_t] dt$ captures the unitary part of the dynamics with  a single Kraus operator $M_{0,0}=\mathbb{1} -iH dt$; the subsequent maps describe the dissipative reservoirs, whose 
 Kraus operators are
\begin{eqnarray}
M_{0,\alpha} &= &\mathbb{1} -\Big(\frac{1}{2}\sum_k L_{k,\alpha}^\dag L_{k,\alpha}\Big)d t  \\
M_{k,\alpha} &= & L_{k,\alpha} \sqrt{d t}, \qquad 1\le k \le K_\alpha.
\end{eqnarray}
Notice that the exact sequence of maps $\E_\alpha$ is immaterial as they all commute to first order in $dt$.
Crucially, each reservoir is assumed to have its own invariant state, $\E_\alpha(\pi^{(\alpha)})=\pi^{(\alpha)}$ (or equivalently $\sum_k {\mathcal D}[L_{k,\alpha}]\pi^{(\alpha)}=0$).
For example, a thermal reservoir at inverse temperature $\beta^{(\alpha)}$ would have the equilibrium Boltzmann density matrix $\pi^{(\alpha)}=e^{\beta^{(\alpha)}(F^{(\alpha)}-H)}$ as its invariant state.
The corresponding Lindblad operators must then induce jumps in that state, $L_{k,\alpha}=\sum_{i,j}m_{ji}^{k,\alpha}|\pi^{(\alpha)}_j\rangle\langle\pi^{(\alpha)}_i|$, to satisfy our generalized detailed balance relation~\eqref{micro-reversibility}.
As a result, the $\{M_{0,\alpha}\}_{\alpha=0}^M$ immediately satisfy the generalized detailed balance relations with $\Delta\Phi_{\pi^{(\alpha)}}=0$, which remarkably does not require the invariant state to commute with the Hamiltonian.

Now, a trajectory for this setup corresponds to a list $\gamma=(n,k^{(\alpha_1)}_1,k^{(\alpha_2)}_2,\dots,k^{(\alpha_N)}_N,m)$, given by the initial and final measurement outcomes, $n$ and $m$, and a set of jumps $k^{(\alpha_l)}_l$ occurring at times $t_l$ in the $\alpha_l$ reservoir.
Notice that only one jump in one of the $M$ reservoirs can happen in any given $dt$, since the probability to observe two jumps is negligible.
The result from \eqref{detailed_ft_concatenation} is  then
\begin{equation}\label{eq:Sigma_multi}
\Sigma(\gamma) =  \sigma(n,m) + \sum_l \Delta\Phi_{\pi^{(\alpha_l)}}(k^{(\alpha_l)}_l).
\end{equation}
This point of view allows us to treat multiple reservoirs at once, such as an engine operating between a hot and cold thermal reservoirs, each represented by a different set of Lindblad operators~\cite{Alicki1979}.
Using the entropic boundary conditions~\eqref{entropy_sigma}, the resulting average entropy production has long been  known from the works of Spohn and Lebowitz~\cite{SpohnLeb1978} and Alicki~\emph{et al.}~\cite{Alicki1979}.

Remarkably, condition \eqref{comml} is fulfilled by almost all known examples of driven Lindblad equations for systems weakly coupled to reservoirs. If the Hamiltonian $H$ of the system is constant, the weak coupling limit results in a Lindblad equation where the operators $L_{\omega},L_{\omega}^\dagger$ are labelled by the Bohr frequencies $\omega$ which are transition frequencies between the levels of the Hamiltonian $H$, i.e, they are of the form $\omega = \omega_i -\omega_j$, for some pair of levels $i,j$ with energies $\epsilon_i=\hbar\omega_i$ and $\epsilon_j=\hbar\omega_j$, respectively \cite{breuer,Rivas:2012wd}. 
These are  ladder operators that lower and raise the energy levels of $H$, obeying the commutation relations: 
\begin{eqnarray}\label{comml2}
[ L_{\omega},H ] = \omega L_{\omega} ~~;~~ {[}L^\dagger_{\omega},H ] = -\omega L^\dagger_{\omega}  
\end{eqnarray}
Their commutator with the logarithm of the stationary density matrix can be written as:
\begin{equation}
\braket{\pi_i|[L_{\omega},\ln\pi]|\pi_j}=\braket{\pi_i|L_{\omega}|\pi_j}\ln\frac{\pi(i)}{\pi(j)}
\end{equation}
For \eqref{comml} to be satisfied it is sufficient that the ratio $\pi(i)/\pi(j)=e^{f(\Delta\epsilon_{ij})}$ is a function of the energy difference $\Delta \epsilon_{ij}=\epsilon_j-\epsilon_i$. In that case
\begin{equation}
[L_{\omega},\ln\pi]=f(\hbar\omega)L_\omega
\end{equation}
and $\Delta\phi_\pi(\omega)=f(\hbar\omega)$. In the case of a single thermal reservoir $f(\epsilon)=\beta\epsilon$, and $\Delta\phi_\pi(\omega)$ is entropy production in the reservoir (heat flow divided by temperature) associated to a transition of frequency $\omega$. 
Furthermore, the Lindblad operators will come in pairs $ \lbrace L_\omega, L_{-\omega} \rbrace $ such that ${\tilde L}_\omega = L_{-\omega} \propto L_\omega^\dagger$, and every jump can be undone.  As a result, the dual process is equivalent to the original process.
This approach was developed for work fluctuations theorems in~\cite{Hekking2013} and heat fluctuations in~\cite{Derezinski2008}.

The preceding arguments can naturally be  extended
to a time-dependent Hamiltonian $H(t)$ and time-dependent Lindblad operators $L_k(t)$, yielding an instantaneous stationary state $\pi(t)$ (or states $\pi^{(\alpha)}(t)$)~\cite{Crooks:2008ji,Crooks2008}.
This is the case when the Hamiltonian $H(t)=H(\lambda_t)$ is driven through the slow change of a collection of external parameter $\lambda_t$, the Lindblad operators become parameterized by the external parameters $L_k(\lambda_t)$, and our generalized detailed balance relation will hold at every time \cite{breuer,Albash:2012dn}. 
They even continue to hold for nearly adiabatic driving~\cite{Suomela2015}.
For fast periodic driving, Floquet theory can be used to derive a Lindblad master equation \cite{Szczygielski:2013wc}.
This theory picks out as a preferred eigenbasis a collection of time-periodic states, or Floquet states, each with a corresponding quasi-energy or Floquet energy.
The collection of Lindblad jump operators $\{L_k\}$ then induce transitions between Floquet eigenstates of the periodic Hamiltonian leading again to  the generalized detailed balance relation \eqref{dual_Lindblad}
with $\Delta \Phi_\pi(k)$ the change in Floquet eigenvalues in the $k$-th jump, which often corresponds to the heat exhausted into the environment \cite{Szczygielski:2013wc,Cuetara2014}.
Finally, our predictions can be used to recover the fluctuation theorems derived for driven Markov dynamics presented in~\cite{Horowitz:2013jd}.

It is remarkable that our fluctuation theorem can yield different results for $\Sigma$, depending on the resolution of the stochastic trajectory.
For instance, in the case of the system in contact with several thermal reservoirs, $\Sigma$ is given by \eqref{eq:Sigma_multi} if the trajectory keeps track of the jumps induced by each reservoir separately.
On the other hand, if the trajectory only gives information about the jumps of the system in the basis where the stationary density matrix of the entire Lindblad equation is diagonal, we have~\eqref{eq:SigmaLindblad}.
Consequently, for the same map one can have both \eqref{eq:SigmaLindblad} and  
\eqref{eq:Sigma_multi}. 
The distinction is the same as the difference between the fluctuation theorem for the entropy production \eqref{eq:Sigma_multi} and the nonadiabatic entropy production~\eqref{eq:SigmaLindblad}~\cite{Esposito:2010to}.

\section{Conclusion}\label{sec:conc}

We have presented a general fluctuation theorem for a large class of completely-positive, trace-preserving quantum maps that verify the generalized detailed balance condition in \eqref{dual_Kraus}.  
From these relations many of the known quantum fluctuation theorems follow naturally.
Included in this family are classical fluctuation theorems for arbitrary stochastic maps, as such maps are special cases of CPTP quantum maps where the dynamics remain diagonal in a particular basis. The theorem exploits the dynamical symmetries of a process and its dual and can be interpreted as a quantum version of the Hatano-Sasa theorem \cite{Hatano:2001uc}.
When specialized to maps induced by thermodynamic reservoirs, our results reproduce the quantum fluctuation theorem for entropy production.

We have extended the notion of the dual process, introduced by Crooks \cite{Crooks:2008ji} and clarified its relationship with the time-reversal process used by Campisi \emph{et al.}~ and Watanabe \emph{et al.}~to derive fluctuation theorems for unitary evolution punctuated by projective measurements \cite{Campisi:2011ka,Watanabe:2014fh} and with the classical dual process used by Esposito and Van den Broeck to split the entropy production into an adiabatic and nonadiabatic contribution \cite{Esposito:2010to}.

For nonunital maps our work should be contrasted with the integral fluctuation theorems presented in~\cite{Albash:2013fq,Rastegin:2014hc,Kafri2012,Goold:2014vx,Goold2015}, which in our notation reads
\begin{equation}
\langle e^{-\sigma}\rangle = \gamma,
\end{equation}
for a process dependent correction factor $\gamma$.
We have seen that by including the nonequilibrium potential, no correction term is necessary 
The resulting $\Sigma$, can then be given a clear interpretation as an entropy production in most setups of physical interest.

Our results also show the peculiarity of unital maps regarding entropy exchange, as already pointed out in ~\cite{Rastegin:2014hc,Rastegin2013,Watanabe:2014fh,Campisi:2011gp,Albash:2013fq}.
The nonequilibrium potential associated to those maps is constant and therefore it does not appear in the fluctuation theorem. The entropy production $\Sigma$ in this case is only given by the boundary terms, suggesting that unital maps can be induced without any entropy exchange between the system and its surrounding. Thermalization at infinite temperature is an obvious example, but decoherence or, equivalently, projective measurements, are relevant examples of unital maps. In all these cases, energy exchange between the system and its surroundings can occur, but this energy exchange does not imply any entropy change in the environment.

\section{Acknowledgements}
This work has been supported by grants ENFASIS (FIS2011-22644) and TerMic (FIS2014-52486-R) from the Spanish Government. 
GMP acknowledges BES-2012-054025. This work also benefited from the COST Action MP1209.

 \bibliography{qft.bib}

\begin{thebibliography}{57}
\expandafter\ifx\csname natexlab\endcsname\relax\def\natexlab#1{#1}\fi
\expandafter\ifx\csname bibnamefont\endcsname\relax
  \def\bibnamefont#1{#1}\fi
\expandafter\ifx\csname bibfnamefont\endcsname\relax
  \def\bibfnamefont#1{#1}\fi
\expandafter\ifx\csname citenamefont\endcsname\relax
  \def\citenamefont#1{#1}\fi
\expandafter\ifx\csname url\endcsname\relax
  \def\url#1{\texttt{#1}}\fi
\expandafter\ifx\csname urlprefix\endcsname\relax\def\urlprefix{URL }\fi
\providecommand{\bibinfo}[2]{#2}
\providecommand{\eprint}[2][]{\url{#2}}

\bibitem[{\citenamefont{Kraus et~al.}(1983)\citenamefont{Kraus, B{\"o}hm,
  Dollard, and Wootters}}]{kraus}
\bibinfo{author}{\bibfnamefont{K.}~\bibnamefont{Kraus}},
  \bibinfo{author}{\bibfnamefont{A.}~\bibnamefont{B{\"o}hm}},
  \bibinfo{author}{\bibfnamefont{J.~D.} \bibnamefont{Dollard}},
  \bibnamefont{and} \bibinfo{author}{\bibfnamefont{W.~H.}
  \bibnamefont{Wootters}}, \emph{\bibinfo{title}{{States, effects, and
  operations : fundamental notions of quantum theory}}}, {Lecture notes in
  physics} (\bibinfo{publisher}{Springer-Verlag}, \bibinfo{address}{Berlin},
  \bibinfo{year}{1983}).

\bibitem[{\citenamefont{Breuer and Petruccione}(2002)}]{breuer}
\bibinfo{author}{\bibfnamefont{H.-P.} \bibnamefont{Breuer}} \bibnamefont{and}
  \bibinfo{author}{\bibfnamefont{F.}~\bibnamefont{Petruccione}},
  \emph{\bibinfo{title}{{The theory of open quantum systems}}}
  (\bibinfo{publisher}{Oxford University Press}, \bibinfo{address}{Oxford},
  \bibinfo{year}{2002}).

\bibitem[{\citenamefont{Yukawa}(2001)}]{Yukawa:2001tf}
\bibinfo{author}{\bibfnamefont{S.}~\bibnamefont{Yukawa}}
  (\bibinfo{year}{2001}), \bibinfo{note}{arXiv:0108421v2}.

\bibitem[{\citenamefont{Sagawa}(2013)}]{Sagawa:2012vd}
\bibinfo{author}{\bibfnamefont{T.}~\bibnamefont{Sagawa}}, in
  \emph{\bibinfo{booktitle}{Lectures on quantum computing, thermodynamics and
  statistical physics}}, edited by
  \bibinfo{editor}{\bibfnamefont{M.}~\bibnamefont{Nakahara}}
  (\bibinfo{publisher}{World Scientific New Jersey}, \bibinfo{year}{2013}),
  vol.~\bibinfo{volume}{8} of \emph{\bibinfo{series}{Kinki University Series on
  Quantum Computing}}.

\bibitem[{\citenamefont{Horowitz and Sagawa}(2014)}]{Horowitz2014b}
\bibinfo{author}{\bibfnamefont{J.~M.} \bibnamefont{Horowitz}} \bibnamefont{and}
  \bibinfo{author}{\bibfnamefont{T.}~\bibnamefont{Sagawa}},
  \bibinfo{journal}{J. Stat. Phys.} \textbf{\bibinfo{volume}{156}},
  \bibinfo{pages}{55} (\bibinfo{year}{2014}).

\bibitem[{\citenamefont{Goold and Modi}(2014)}]{Goold:2014vx}
\bibinfo{author}{\bibfnamefont{J.}~\bibnamefont{Goold}} \bibnamefont{and}
  \bibinfo{author}{\bibfnamefont{K.}~\bibnamefont{Modi}},
  \bibinfo{journal}{arXiv}  (\bibinfo{year}{2014}), \eprint{1407.4618v1}.

\bibitem[{\citenamefont{Binder et~al.}(2015)\citenamefont{Binder,
  Vinjanampathy, Modi, and Goold}}]{Binder2015}
\bibinfo{author}{\bibfnamefont{F.}~\bibnamefont{Binder}},
  \bibinfo{author}{\bibfnamefont{S.}~\bibnamefont{Vinjanampathy}},
  \bibinfo{author}{\bibfnamefont{K.}~\bibnamefont{Modi}}, \bibnamefont{and}
  \bibinfo{author}{\bibfnamefont{J.}~\bibnamefont{Goold}},
  \bibinfo{journal}{Phys. Rev. E} \textbf{\bibinfo{volume}{91}},
  \bibinfo{pages}{032119} (\bibinfo{year}{2015}).

\bibitem[{\citenamefont{Campisi
  et~al.}(2011{\natexlab{a}})\citenamefont{Campisi, H{\"a}nggi, and
  Talkner}}]{Campisi:2011ka}
\bibinfo{author}{\bibfnamefont{M.}~\bibnamefont{Campisi}},
  \bibinfo{author}{\bibfnamefont{P.}~\bibnamefont{H{\"a}nggi}},
  \bibnamefont{and} \bibinfo{author}{\bibfnamefont{P.}~\bibnamefont{Talkner}},
  \bibinfo{journal}{Rev Mod Phys} \textbf{\bibinfo{volume}{83}},
  \bibinfo{pages}{771} (\bibinfo{year}{2011}{\natexlab{a}}).

\bibitem[{\citenamefont{Esposito et~al.}(2009)\citenamefont{Esposito, Harbola,
  and Mukamel}}]{Esposito2009}
\bibinfo{author}{\bibfnamefont{M.}~\bibnamefont{Esposito}},
  \bibinfo{author}{\bibfnamefont{U.}~\bibnamefont{Harbola}}, \bibnamefont{and}
  \bibinfo{author}{\bibfnamefont{S.}~\bibnamefont{Mukamel}},
  \bibinfo{journal}{Rev. Mod. Phys.} \textbf{\bibinfo{volume}{81}},
  \bibinfo{pages}{1665} (\bibinfo{year}{2009}).

\bibitem[{\citenamefont{Deffner and Lutz}(2011)}]{Deffner2011}
\bibinfo{author}{\bibfnamefont{S.}~\bibnamefont{Deffner}} \bibnamefont{and}
  \bibinfo{author}{\bibfnamefont{E.}~\bibnamefont{Lutz}},
  \bibinfo{journal}{Phys. Rev. Lett.} \textbf{\bibinfo{volume}{107}}
  (\bibinfo{year}{2011}).

\bibitem[{\citenamefont{Campisi et~al.}(2010)\citenamefont{Campisi, Talkner,
  and H{\"a}nggi}}]{Campisi:2010kx}
\bibinfo{author}{\bibfnamefont{M.}~\bibnamefont{Campisi}},
  \bibinfo{author}{\bibfnamefont{P.}~\bibnamefont{Talkner}}, \bibnamefont{and}
  \bibinfo{author}{\bibfnamefont{P.}~\bibnamefont{H{\"a}nggi}},
  \bibinfo{journal}{Phys. Rev. Lett.} \textbf{\bibinfo{volume}{105}},
  \bibinfo{pages}{140601} (\bibinfo{year}{2010}).

\bibitem[{\citenamefont{Campisi
  et~al.}(2011{\natexlab{b}})\citenamefont{Campisi, Talkner, and
  H{\"a}nggi}}]{Campisi:2011gp}
\bibinfo{author}{\bibfnamefont{M.}~\bibnamefont{Campisi}},
  \bibinfo{author}{\bibfnamefont{P.}~\bibnamefont{Talkner}}, \bibnamefont{and}
  \bibinfo{author}{\bibfnamefont{P.}~\bibnamefont{H{\"a}nggi}},
  \bibinfo{journal}{Phys Rev E} \textbf{\bibinfo{volume}{83}},
  \bibinfo{pages}{041114} (\bibinfo{year}{2011}{\natexlab{b}}).

\bibitem[{\citenamefont{Watanabe et~al.}(2014)\citenamefont{Watanabe,
  Venkatesh, Talkner, Campisi, and H{\"a}nggi}}]{Watanabe:2014fh}
\bibinfo{author}{\bibfnamefont{G.}~\bibnamefont{Watanabe}},
  \bibinfo{author}{\bibfnamefont{B.~P.} \bibnamefont{Venkatesh}},
  \bibinfo{author}{\bibfnamefont{P.}~\bibnamefont{Talkner}},
  \bibinfo{author}{\bibfnamefont{M.}~\bibnamefont{Campisi}}, \bibnamefont{and}
  \bibinfo{author}{\bibfnamefont{P.}~\bibnamefont{H{\"a}nggi}},
  \bibinfo{journal}{Phys Rev E} \textbf{\bibinfo{volume}{89}},
  \bibinfo{pages}{032114} (\bibinfo{year}{2014}).

\bibitem[{\citenamefont{Crooks}(2008{\natexlab{a}})}]{Crooks:2008ji}
\bibinfo{author}{\bibfnamefont{G.~E.} \bibnamefont{Crooks}},
  \bibinfo{journal}{Phys. Rev. A} \textbf{\bibinfo{volume}{77}},
  \bibinfo{pages}{034101} (\bibinfo{year}{2008}{\natexlab{a}}).

\bibitem[{\citenamefont{Horowitz and Parrondo}(2013)}]{Horowitz:2013jd}
\bibinfo{author}{\bibfnamefont{J.~M.} \bibnamefont{Horowitz}} \bibnamefont{and}
  \bibinfo{author}{\bibfnamefont{J.~M.~R.} \bibnamefont{Parrondo}},
  \bibinfo{journal}{New J Phys} \textbf{\bibinfo{volume}{15}},
  \bibinfo{pages}{085028} (\bibinfo{year}{2013}).

\bibitem[{\citenamefont{Chetritie and Mallick}(2012)}]{Chetrite2012}
\bibinfo{author}{\bibfnamefont{R.}~\bibnamefont{Chetritie}} \bibnamefont{and}
  \bibinfo{author}{\bibfnamefont{K.}~\bibnamefont{Mallick}},
  \bibinfo{journal}{J. Stat. Phys.} \textbf{\bibinfo{volume}{148}},
  \bibinfo{pages}{480} (\bibinfo{year}{2012}).

\bibitem[{\citenamefont{Rastegin}(2013)}]{Rastegin2013}
\bibinfo{author}{\bibfnamefont{A.~E.} \bibnamefont{Rastegin}},
  \bibinfo{journal}{J. Stat. Mech.: Theor. Exp.} p. \bibinfo{pages}{P06016}
  (\bibinfo{year}{2013}).

\bibitem[{\citenamefont{Rastegin and {\.Z}yczkowski}(2014)}]{Rastegin:2014hc}
\bibinfo{author}{\bibfnamefont{A.~E.} \bibnamefont{Rastegin}} \bibnamefont{and}
  \bibinfo{author}{\bibfnamefont{K.}~\bibnamefont{{\.Z}yczkowski}},
  \bibinfo{journal}{Physical Review E} \textbf{\bibinfo{volume}{89}},
  \bibinfo{pages}{012127} (\bibinfo{year}{2014}).

\bibitem[{\citenamefont{Albash et~al.}(2013)\citenamefont{Albash, Lidar,
  Marvian, and Zanardi}}]{Albash:2013fq}
\bibinfo{author}{\bibfnamefont{T.}~\bibnamefont{Albash}},
  \bibinfo{author}{\bibfnamefont{D.~A.} \bibnamefont{Lidar}},
  \bibinfo{author}{\bibfnamefont{M.}~\bibnamefont{Marvian}}, \bibnamefont{and}
  \bibinfo{author}{\bibfnamefont{P.}~\bibnamefont{Zanardi}},
  \bibinfo{journal}{Physical Review E} \textbf{\bibinfo{volume}{88}},
  \bibinfo{pages}{032146} (\bibinfo{year}{2013}).

\bibitem[{\citenamefont{Kafri and Deffner}(2012)}]{Kafri2012}
\bibinfo{author}{\bibfnamefont{D.}~\bibnamefont{Kafri}} \bibnamefont{and}
  \bibinfo{author}{\bibfnamefont{S.}~\bibnamefont{Deffner}},
  \bibinfo{journal}{Phys. Rev. A} \textbf{\bibinfo{volume}{86}},
  \bibinfo{pages}{044302} (\bibinfo{year}{2012}).

\bibitem[{\citenamefont{Goold et~al.}(2015)\citenamefont{Goold, Paternostro,
  and Modi}}]{Goold2015}
\bibinfo{author}{\bibfnamefont{J.}~\bibnamefont{Goold}},
  \bibinfo{author}{\bibfnamefont{M.}~\bibnamefont{Paternostro}},
  \bibnamefont{and} \bibinfo{author}{\bibfnamefont{K.}~\bibnamefont{Modi}},
  \bibinfo{journal}{Phys. Rev. Lett.} \textbf{\bibinfo{volume}{114}},
  \bibinfo{pages}{060602} (\bibinfo{year}{2015}).

\bibitem[{\citenamefont{Scully et~al.}(2003)\citenamefont{Scully, Zubairy,
  Agarwal, and Walther}}]{Scully2003}
\bibinfo{author}{\bibfnamefont{M.~O.} \bibnamefont{Scully}},
  \bibinfo{author}{\bibfnamefont{M.~S.} \bibnamefont{Zubairy}},
  \bibinfo{author}{\bibfnamefont{G.~S.} \bibnamefont{Agarwal}},
  \bibnamefont{and} \bibinfo{author}{\bibfnamefont{H.}~\bibnamefont{Walther}},
  \bibinfo{journal}{Science} \textbf{\bibinfo{volume}{299}},
  \bibinfo{pages}{862} (\bibinfo{year}{2003}).

\bibitem[{\citenamefont{Lutz and Dillenschneider}(2009)}]{Lutz2009}
\bibinfo{author}{\bibfnamefont{E.}~\bibnamefont{Lutz}} \bibnamefont{and}
  \bibinfo{author}{\bibfnamefont{R.}~\bibnamefont{Dillenschneider}},
  \bibinfo{journal}{Europhys. Lett.} \textbf{\bibinfo{volume}{88}},
  \bibinfo{pages}{50003} (\bibinfo{year}{2009}).

\bibitem[{\citenamefont{Wiseman and Milburn}(2010)}]{wiseman}
\bibinfo{author}{\bibfnamefont{H.~M.} \bibnamefont{Wiseman}} \bibnamefont{and}
  \bibinfo{author}{\bibfnamefont{G.~J.} \bibnamefont{Milburn}},
  \emph{\bibinfo{title}{{Quantum measurement and control}}}
  (\bibinfo{publisher}{Cambridge University Press},
  \bibinfo{address}{Cambridge, UK}, \bibinfo{year}{2010}).

\bibitem[{\citenamefont{Haake}(2010)}]{Haake}
\bibinfo{author}{\bibfnamefont{F.}~\bibnamefont{Haake}},
  \emph{\bibinfo{title}{{Quantum signatures of chaos}}}, {Springer series in
  synergetics,} (\bibinfo{publisher}{Springer}, \bibinfo{address}{Berlin},
  \bibinfo{year}{2010}), \bibinfo{edition}{3rd} ed.

\bibitem[{\citenamefont{Andrieux and Gaspard}(2008)}]{Andrieux:2008em}
\bibinfo{author}{\bibfnamefont{D.}~\bibnamefont{Andrieux}} \bibnamefont{and}
  \bibinfo{author}{\bibfnamefont{P.}~\bibnamefont{Gaspard}},
  \bibinfo{journal}{Phys. Rev. Lett.} \textbf{\bibinfo{volume}{100}},
  \bibinfo{pages}{230404} (\bibinfo{year}{2008}).

\bibitem[{\citenamefont{Maes}(1999)}]{Maes:1999kn}
\bibinfo{author}{\bibfnamefont{C.}~\bibnamefont{Maes}}, \bibinfo{journal}{J
  Stat Phys} \textbf{\bibinfo{volume}{95}}, \bibinfo{pages}{367}
  (\bibinfo{year}{1999}).

\bibitem[{\citenamefont{Hurtado et~al.}(2011)\citenamefont{Hurtado,
  Perez-Espigares, del Pozo, and Garrido}}]{Hurtado:2011gm}
\bibinfo{author}{\bibfnamefont{P.~I.} \bibnamefont{Hurtado}},
  \bibinfo{author}{\bibfnamefont{C.}~\bibnamefont{Perez-Espigares}},
  \bibinfo{author}{\bibfnamefont{J.~J.} \bibnamefont{del Pozo}},
  \bibnamefont{and} \bibinfo{author}{\bibfnamefont{P.~L.}
  \bibnamefont{Garrido}}, \bibinfo{journal}{PNAS}
  \textbf{\bibinfo{volume}{108}}, \bibinfo{pages}{7704} (\bibinfo{year}{2011}).

\bibitem[{\citenamefont{Lacoste and Gaspard}(2014)}]{Lacoste:2014kh}
\bibinfo{author}{\bibfnamefont{D.}~\bibnamefont{Lacoste}} \bibnamefont{and}
  \bibinfo{author}{\bibfnamefont{P.}~\bibnamefont{Gaspard}},
  \bibinfo{journal}{Phys. Rev. Lett.} \textbf{\bibinfo{volume}{113}},
  \bibinfo{pages}{240602} (\bibinfo{year}{2014}).

\bibitem[{\citenamefont{Hatano and Sasa}(2001)}]{Hatano:2001uc}
\bibinfo{author}{\bibfnamefont{T.}~\bibnamefont{Hatano}} \bibnamefont{and}
  \bibinfo{author}{\bibfnamefont{S.-i.} \bibnamefont{Sasa}},
  \bibinfo{journal}{Phys. Rev. Lett.} \textbf{\bibinfo{volume}{86}},
  \bibinfo{pages}{3463} (\bibinfo{year}{2001}).

\bibitem[{\citenamefont{Horowitz}(2012)}]{Horowitz:2012gn}
\bibinfo{author}{\bibfnamefont{J.~M.} \bibnamefont{Horowitz}},
  \bibinfo{journal}{Phys Rev E} \textbf{\bibinfo{volume}{85}},
  \bibinfo{pages}{031110} (\bibinfo{year}{2012}).

\bibitem[{\citenamefont{Alicki et~al.}(2006)\citenamefont{Alicki, Lidar, and
  Zanardi}}]{Alicki:2006tr}
\bibinfo{author}{\bibfnamefont{R.}~\bibnamefont{Alicki}},
  \bibinfo{author}{\bibfnamefont{D.~A.} \bibnamefont{Lidar}}, \bibnamefont{and}
  \bibinfo{author}{\bibfnamefont{P.}~\bibnamefont{Zanardi}},
  \bibinfo{journal}{Phys. Rev. A} \textbf{\bibinfo{volume}{73}},
  \bibinfo{pages}{052311} (\bibinfo{year}{2006}).

\bibitem[{\citenamefont{Szczygielski et~al.}(2013)\citenamefont{Szczygielski,
  Gelbwaser-Klimovsky, and Alicki}}]{Szczygielski:2013wc}
\bibinfo{author}{\bibfnamefont{K.}~\bibnamefont{Szczygielski}},
  \bibinfo{author}{\bibfnamefont{D.}~\bibnamefont{Gelbwaser-Klimovsky}},
  \bibnamefont{and} \bibinfo{author}{\bibfnamefont{R.}~\bibnamefont{Alicki}},
  \bibinfo{journal}{Phys Rev E} \textbf{\bibinfo{volume}{87}},
  \bibinfo{pages}{012120} (\bibinfo{year}{2013}).

\bibitem[{\citenamefont{Rivas and Huelga}(2012)}]{Rivas:2012wd}
\bibinfo{author}{\bibfnamefont{A.}~\bibnamefont{Rivas}} \bibnamefont{and}
  \bibinfo{author}{\bibfnamefont{S.~F.} \bibnamefont{Huelga}},
  \emph{\bibinfo{title}{{Open Quantum Systems : An Introduction}}}
  (\bibinfo{publisher}{Springer Berlin Heidelberg}, \bibinfo{address}{Berlin,
  Heidelberg}, \bibinfo{year}{2012}).

\bibitem[{\citenamefont{Kawai et~al.}(2007)\citenamefont{Kawai, Parrondo, and
  den Broeck}}]{Kawai:2007kc}
\bibinfo{author}{\bibfnamefont{R.}~\bibnamefont{Kawai}},
  \bibinfo{author}{\bibfnamefont{J.}~\bibnamefont{Parrondo}}, \bibnamefont{and}
  \bibinfo{author}{\bibfnamefont{C.}~\bibnamefont{den Broeck}},
  \bibinfo{journal}{Phys. Rev. Lett.} \textbf{\bibinfo{volume}{98}}
  (\bibinfo{year}{2007}).

\bibitem[{\citenamefont{Esposito and {Van den
  Broeck}}(2010{\natexlab{a}})}]{Esposito:2010jf}
\bibinfo{author}{\bibfnamefont{M.}~\bibnamefont{Esposito}} \bibnamefont{and}
  \bibinfo{author}{\bibfnamefont{C.}~\bibnamefont{{Van den Broeck}}},
  \bibinfo{journal}{Phys Rev E} \textbf{\bibinfo{volume}{82}},
  \bibinfo{pages}{011143} (\bibinfo{year}{2010}{\natexlab{a}}).

\bibitem[{\citenamefont{{Van den Broeck} and Esposito}(2010)}]{Esposito:2010vu}
\bibinfo{author}{\bibfnamefont{C.}~\bibnamefont{{Van den Broeck}}}
  \bibnamefont{and} \bibinfo{author}{\bibfnamefont{M.}~\bibnamefont{Esposito}},
  \bibinfo{journal}{Phys Rev E} \textbf{\bibinfo{volume}{82}},
  \bibinfo{pages}{011144} (\bibinfo{year}{2010}).

\bibitem[{\citenamefont{Esposito and {Van den
  Broeck}}(2010{\natexlab{b}})}]{Esposito:2010to}
\bibinfo{author}{\bibfnamefont{M.}~\bibnamefont{Esposito}} \bibnamefont{and}
  \bibinfo{author}{\bibfnamefont{C.}~\bibnamefont{{Van den Broeck}}},
  \bibinfo{journal}{Phys. Rev. Lett.} \textbf{\bibinfo{volume}{104}}
  (\bibinfo{year}{2010}{\natexlab{b}}).

\bibitem[{\citenamefont{Chernyak et~al.}(2006)\citenamefont{Chernyak, Chertkov,
  and Jarzynsk}}]{Chernyak2006b}
\bibinfo{author}{\bibfnamefont{V.~Y.} \bibnamefont{Chernyak}},
  \bibinfo{author}{\bibfnamefont{M.}~\bibnamefont{Chertkov}}, \bibnamefont{and}
  \bibinfo{author}{\bibfnamefont{C.}~\bibnamefont{Jarzynsk}},
  \bibinfo{journal}{J. Stat. Mech.: Theor. Exp.} p. \bibinfo{pages}{P08001}
  (\bibinfo{year}{2006}).

\bibitem[{\citenamefont{Speck and Seifert}(2005)}]{Speck2005b}
\bibinfo{author}{\bibfnamefont{T.}~\bibnamefont{Speck}} \bibnamefont{and}
  \bibinfo{author}{\bibfnamefont{U.}~\bibnamefont{Seifert}},
  \bibinfo{journal}{J. Phys. A: Math. Gen.} \textbf{\bibinfo{volume}{38}},
  \bibinfo{pages}{L581} (\bibinfo{year}{2005}).

\bibitem[{\citenamefont{Seifert}(2005)}]{Seifert:2005fu}
\bibinfo{author}{\bibfnamefont{U.}~\bibnamefont{Seifert}},
  \bibinfo{journal}{Phys. Rev. Lett.} \textbf{\bibinfo{volume}{95}},
  \bibinfo{pages}{040602} (\bibinfo{year}{2005}).

\bibitem[{\citenamefont{Monnai}(2005)}]{Monnai2005}
\bibinfo{author}{\bibfnamefont{T.}~\bibnamefont{Monnai}},
  \bibinfo{journal}{Phys. Rev. E} \textbf{\bibinfo{volume}{72}},
  \bibinfo{pages}{027102} (\bibinfo{year}{2005}).

\bibitem[{\citenamefont{Batalh{\~a}o et~al.}(2014)\citenamefont{Batalh{\~a}o,
  Souza, Mazzola, Auccaise, Sarthour, Oliveira, Goold, {De Chiara},
  Paternostro, and Serra}}]{Batalhao:2014ta}
\bibinfo{author}{\bibfnamefont{T.~B.} \bibnamefont{Batalh{\~a}o}},
  \bibinfo{author}{\bibfnamefont{A.~M.} \bibnamefont{Souza}},
  \bibinfo{author}{\bibfnamefont{L.}~\bibnamefont{Mazzola}},
  \bibinfo{author}{\bibfnamefont{R.}~\bibnamefont{Auccaise}},
  \bibinfo{author}{\bibfnamefont{R.~S.} \bibnamefont{Sarthour}},
  \bibinfo{author}{\bibfnamefont{I.~S.} \bibnamefont{Oliveira}},
  \bibinfo{author}{\bibfnamefont{J.}~\bibnamefont{Goold}},
  \bibinfo{author}{\bibfnamefont{G.}~\bibnamefont{{De Chiara}}},
  \bibinfo{author}{\bibfnamefont{M.}~\bibnamefont{Paternostro}},
  \bibnamefont{and} \bibinfo{author}{\bibfnamefont{R.~M.} \bibnamefont{Serra}},
  \bibinfo{journal}{Phys. Rev. Lett.} \textbf{\bibinfo{volume}{113}},
  \bibinfo{pages}{140601} (\bibinfo{year}{2014}).

\bibitem[{\citenamefont{Bentsoon and Zyczkowski}(2006)}]{Bengtsson}
\bibinfo{author}{\bibfnamefont{I.}~\bibnamefont{Bentsoon}} \bibnamefont{and}
  \bibinfo{author}{\bibfnamefont{K.}~\bibnamefont{Zyczkowski}},
  \emph{\bibinfo{title}{{Geometry of Quantum States: An Introduction to Quantum
  Entanglement}}} (\bibinfo{publisher}{University Press, Cambridge},
  \bibinfo{year}{2006}).

\bibitem[{\citenamefont{Callens et~al.}(2004)\citenamefont{Callens, {De Roeck},
  Jacobs, Maes, and Neto\v{c}n{\'y}}}]{Callens2004}
\bibinfo{author}{\bibfnamefont{I.}~\bibnamefont{Callens}},
  \bibinfo{author}{\bibfnamefont{W.}~\bibnamefont{{De Roeck}}},
  \bibinfo{author}{\bibfnamefont{T.}~\bibnamefont{Jacobs}},
  \bibinfo{author}{\bibfnamefont{C.}~\bibnamefont{Maes}}, \bibnamefont{and}
  \bibinfo{author}{\bibfnamefont{K.}~\bibnamefont{Neto\v{c}n{\'y}}},
  \bibinfo{journal}{Physica D} \textbf{\bibinfo{volume}{187}},
  \bibinfo{pages}{383} (\bibinfo{year}{2004}).

\bibitem[{\citenamefont{{De Roeck} and Maes}(2004)}]{DeRoeck2004}
\bibinfo{author}{\bibfnamefont{W.}~\bibnamefont{{De Roeck}}} \bibnamefont{and}
  \bibinfo{author}{\bibfnamefont{C.}~\bibnamefont{Maes}},
  \bibinfo{journal}{Phys. Rev. E} \textbf{\bibinfo{volume}{69}},
  \bibinfo{pages}{026115} (\bibinfo{year}{2004}).

\bibitem[{\citenamefont{Faist et~al.}(2015)\citenamefont{Faist, Oppenheim, and
  Renner}}]{Faist2015}
\bibinfo{author}{\bibfnamefont{P.}~\bibnamefont{Faist}},
  \bibinfo{author}{\bibfnamefont{J.}~\bibnamefont{Oppenheim}},
  \bibnamefont{and} \bibinfo{author}{\bibfnamefont{R.}~\bibnamefont{Renner}},
  \bibinfo{journal}{New J Phys} \textbf{\bibinfo{volume}{17}}
  (\bibinfo{year}{2015}).

\bibitem[{\citenamefont{Fagnola and Umanit{\`a}}(2007)}]{Fagnola:2007hj}
\bibinfo{author}{\bibfnamefont{F.}~\bibnamefont{Fagnola}} \bibnamefont{and}
  \bibinfo{author}{\bibfnamefont{V.}~\bibnamefont{Umanit{\`a}}},
  \bibinfo{journal}{Infin. Dimens. Anal. Quantum. Probab. Relat. Top.}
  \textbf{\bibinfo{volume}{10}}, \bibinfo{pages}{335} (\bibinfo{year}{2007}).

\bibitem[{\citenamefont{Spohn}(1978)}]{Spohn1978}
\bibinfo{author}{\bibfnamefont{H.}~\bibnamefont{Spohn}}, \bibinfo{journal}{J.
  Math. Phys.} \textbf{\bibinfo{volume}{19}}, \bibinfo{pages}{1227}
  (\bibinfo{year}{1978}).

\bibitem[{\citenamefont{Alicki}(1979)}]{Alicki1979}
\bibinfo{author}{\bibfnamefont{R.}~\bibnamefont{Alicki}}, \bibinfo{journal}{J.
  Phys. A} \textbf{\bibinfo{volume}{12}}, \bibinfo{pages}{L103}
  (\bibinfo{year}{1979}).

\bibitem[{\citenamefont{Spohn and Lebowitz}(1978)}]{SpohnLeb1978}
\bibinfo{author}{\bibfnamefont{H.}~\bibnamefont{Spohn}} \bibnamefont{and}
  \bibinfo{author}{\bibfnamefont{J.~L.} \bibnamefont{Lebowitz}}, in
  \emph{\bibinfo{booktitle}{{Advances in Chemical Physics: For Ilya
  Prigogine}}}, edited by \bibinfo{editor}{\bibfnamefont{S.~A.}
  \bibnamefont{Rice}} (\bibinfo{publisher}{John Wiley \& Sons, Hoboken, NJ},
  \bibinfo{year}{1978}), vol.~\bibinfo{volume}{38}.

\bibitem[{\citenamefont{Hekking and Pekola}(2013)}]{Hekking2013}
\bibinfo{author}{\bibfnamefont{F.~W.~J.} \bibnamefont{Hekking}}
  \bibnamefont{and} \bibinfo{author}{\bibfnamefont{J.~P.}
  \bibnamefont{Pekola}}, \bibinfo{journal}{Phys. Rev. Lett.}
  \textbf{\bibinfo{volume}{111}}, \bibinfo{pages}{093602}
  (\bibinfo{year}{2013}).

\bibitem[{\citenamefont{Derezi{\'n}ski
  et~al.}(2008)\citenamefont{Derezi{\'n}ski, {De Roeck}, and
  Maes}}]{Derezinski2008}
\bibinfo{author}{\bibfnamefont{J.}~\bibnamefont{Derezi{\'n}ski}},
  \bibinfo{author}{\bibfnamefont{W.}~\bibnamefont{{De Roeck}}},
  \bibnamefont{and} \bibinfo{author}{\bibfnamefont{C.}~\bibnamefont{Maes}},
  \bibinfo{journal}{J. Stat. Phys.} \textbf{\bibinfo{volume}{131}},
  \bibinfo{pages}{341} (\bibinfo{year}{2008}).

\bibitem[{\citenamefont{Crooks}(2008{\natexlab{b}})}]{Crooks2008}
\bibinfo{author}{\bibfnamefont{G.~E.} \bibnamefont{Crooks}},
  \bibinfo{journal}{J. Stat. Mech.: Theor. Exp.} \textbf{\bibinfo{volume}{10}},
  \bibinfo{pages}{P10023} (\bibinfo{year}{2008}{\natexlab{b}}).

\bibitem[{\citenamefont{Albash et~al.}(2012)\citenamefont{Albash, Boixo, Lidar,
  and Zanardi}}]{Albash:2012dn}
\bibinfo{author}{\bibfnamefont{T.}~\bibnamefont{Albash}},
  \bibinfo{author}{\bibfnamefont{S.}~\bibnamefont{Boixo}},
  \bibinfo{author}{\bibfnamefont{D.~A.} \bibnamefont{Lidar}}, \bibnamefont{and}
  \bibinfo{author}{\bibfnamefont{P.}~\bibnamefont{Zanardi}},
  \bibinfo{journal}{New J Phys} \textbf{\bibinfo{volume}{14}},
  \bibinfo{pages}{123016} (\bibinfo{year}{2012}).

\bibitem[{\citenamefont{Suomela et~al.}(2015)\citenamefont{Suomela, Salmilehto,
  Savenko, Ala-Nissila, and M{\"o}tt{\"o}nen}}]{Suomela2015}
\bibinfo{author}{\bibfnamefont{S.}~\bibnamefont{Suomela}},
  \bibinfo{author}{\bibfnamefont{J.}~\bibnamefont{Salmilehto}},
  \bibinfo{author}{\bibfnamefont{I.~G.} \bibnamefont{Savenko}},
  \bibinfo{author}{\bibfnamefont{T.}~\bibnamefont{Ala-Nissila}},
  \bibnamefont{and}
  \bibinfo{author}{\bibfnamefont{M.}~\bibnamefont{M{\"o}tt{\"o}nen}},
  \bibinfo{journal}{Phys. Rev. E} \textbf{\bibinfo{volume}{91}},
  \bibinfo{pages}{022126} (\bibinfo{year}{2015}).

\bibitem[{\citenamefont{Cuetara et~al.}(2015)\citenamefont{Cuetara, Engel, and
  Esposito}}]{Cuetara2014}
\bibinfo{author}{\bibfnamefont{G.~B.} \bibnamefont{Cuetara}},
  \bibinfo{author}{\bibfnamefont{A.}~\bibnamefont{Engel}}, \bibnamefont{and}
  \bibinfo{author}{\bibfnamefont{M.}~\bibnamefont{Esposito}},
  \bibinfo{journal}{New J Phys} \textbf{\bibinfo{volume}{17}},
  \bibinfo{pages}{055002} (\bibinfo{year}{2015}).

\end{thebibliography}
 
\end{document}